\documentstyle[aaspp4]{article}
\begin{document}
\title{ A Compact Fireball Model of Gamma Ray Bursts}
\author{David Eichler \& Amir Levinson}
\begin{abstract}
It is proposed that the  gamma ray burst photons near the peak of
the spectrum at several hundred KeV are produced on very compact
scales, where photon production is limited by blackbody effects
and/or the requirement of energetic quanta ($E>2m_e c^2$) for
efficient further production. The fast variation of order
milliseconds in the time profile is then a natural expectation,
given  the other observed GRB parameters. Analytic calculations
are presented to show that the escape of  non-thermal, energetic
gamma rays
 can emerge within a second of the thermal photons from a gammasphere of
below $10^{12}$ cm.
The minimum asymptotic bulk Lorentz factor in this model is found to be
of order several hundred if the photosphere is of order $3 \times 10^{11}$
cm and greater  for larger or smaller photospheric radii. It is suggested that
prompt UHE gamma rays might provide a new constraint on the asymptotic Lorentz factor of the outflow.
\end{abstract}

\section{Introduction}

Gamma ray burst spectra  typically peak at several hundred KeV,
and frequently (though not always)  have an extended non-thermal
tail that contains a significant fraction of the total energy.
Although the high energy side of the peak is hard to characterize
with reliable generalizations, classical GRB's seem to typically
peak above 100 KeV; a significantly  softer peak is a sign that
the GRB is in a separate class of "soft repeaters". Though this
has been suggested to be a selection effect, we believe it to be
real. The total luminosity in soft X-rays is typically below  the
luminosity above 100 KeV by an order of magnitude or more. When
one considers that, realistically, an emitting surface will have
some dispersion in its local temperature  (e.g.  a hot spot is
likely to have  cool edges, relativistic beaming from material
moving directly at the observer is likely to be mixed with
contributions from fluid with a sideways component to its motion
etc.) the X-ray paucity seems all the more  clean and
significant.

We believe that the characteristic peak energy of about 200 KeV is not
merely a chance value or selection effect, but is rather telling us
something important about GRB's.
In this paper we construct a model that, by design, yields this
characteristic peak
energy naturally. The basic assertion is that the gamma rays near the peak
are emitted from a compact region of radius R not much greater than
about $10^{9}\Gamma$cm, where $\Gamma$ is the bulk Lorentz factor of the
expanding fireball.  Earlier discussions and variations of this basic idea
include those by Eichler (1994) and Thompson (1997).
Energization of the pair plasma at this radius
could come from internal shocks (e.g., Eichler 1994; Rees \& Meszaros 1994;
Sari \& Piran 1997)
but could also come from, say, collimation by surrounding baryonic material.
The possibility of collimation may be motivated by  the huge fluence
from recently discovered GRB's such as GRB 990123. Interaction with the
collimating material could produce strong internal shock, or  could
proceed via radiative viscosity with the wall of the collimating material.

In the latter process, dissipation (i.e. creation of additional quanta)
would proceed very efficiently when the average photon energy in the
observer frame approached $m_ec^2$, for then large angle scattering
  by the walls of the collimating material of photons
back into the jet, where they could appear in the local frame to be
blueshifted to even higher energy,
would allow creation of additional pairs.
Once the average energy per quantum in the observer frame goes far enough
below $m_ec^2$, this dissipation mechanism would taper off. This is true
of
any  dissipation mechanism that proceeds by pair creation.  The point is
that, if the clustering of GRB peaks  near or just below $m_ec^2$ is to
be regarded as a signature of dissipation by pair production, then  such
dissipation would probably occur  at modest $\Gamma$, or else the product
of large $\Gamma$ with a small fireball frame energy per quantum
consistently yielding a given value for the observer's frame would seem a
puzzling  coincidence.  If the dissipation is to take place at modest
$\Gamma$, we argue that this probably indicates compact scales, before
the fireball accelerates to its asymptotic value of $\Gamma$, which
according
to several considerations is quite large ($\ge 300$).

At compact scales,
thermalization is a strong possibility, and sets a convenient reference
point for the analysis, but
there is no implication that non-thermal effects are absent, or that the
photon spectrum need look thermal on either side of the peak. Our view is
that the non-thermal component in GRB's is not, insofar as can be
currently established by observations, as remarkably consistent
from burst to burst as is the spectral peak. Thus, non-thermal
acceleration might proceed  downstream of the photosphere, as in
optically thin synchrotron models of bursts with only the constraint that
the non-thermal component never dwarf the thermal component (in total
photons produced). The
rationale for  why both components might be of the same order of magnitude
is given in section 3, where it is shown that the energy budget of the
burst released in the internal shocks of an unsteady flow
might be distributed logarithmically over a broad range of radii.

It is also possible that  direct shock energization of photons (Blandford
and Payne, 1982) can create a non-thermal photon spectrum up to 100 KeV
in the frame of the fluid just below the photosphere by a shock that
passes through the photosphere of a GRB (Eichler 1994). The photons
would escape before they are
thermalized or  saturatedly Comptonized.  We will argue
in this paper that the material at the photosphere could be moving with a
bulk Lorentz factor $\Gamma$ that exceeds 300, depending on the extent to
which
non-thermal processes maintain a pair population well above the Boltzmann
equilibrium value. Thus the high energy cutoff of the non-thermal spectrum
could be above 30 MeV in the frame of the observer. The total photon
number would have been established to within a factor of order unity well
within the photosphere (at an optical depth of order $1/\alpha$ if by
bremsstrahlung). On the other hand, the photon number density could
change by a factor of order  unity over the hydrodynamical timescale if
the expansion, assumed to be unsteady, deviates from adiabaticity by a
substantial margin. In this case, self absorption, which would keep the
photon number steady in a static environment, fails to do so in the course
of the expansion. As the Comptonization timescale just below the
photosphere is at least as long as the hydrodynamical timescale, the
spectrum of soft photons  is not fully saturated by Comptonization.

Since the photospheric radius in our model is considerably smaller
than in some other models, which by design recognize the restrictions
on R and $\Gamma$  implied by the escape of energetic, non-thermal
gamma rays, we redo the calculation of this restriction in a transparent
manner and show that it is consistent with the other restrictions that we
invoke here. These calculations show that prompt very energetic gamma rays
are possible, even prompt TeV ones, depending on the strength of the
magnetic field.  Conceivably, detection and time resolution of very
energetic gamma
rays in future experiments could set a useful diagnostic of the field
strength and bulk Lorentz factor.

While  the constraint of X-ray paucity might
not be rigorous, and reconcilable with a "just-so" selection of
parameters, we note that, historically, it was  taken seriously
(Imamura \& Epstein 1987) before the cosmological GRB picture was
made popular by the all sky isotropy established
by BATSE observations.

To avoid confusion,  we define several distinct time intervals:
The elapsed time in the frame of the fireball as the
fireball evolves through radius R is called $\Delta t'$ and is equal to
$R/c\Gamma$. The characteristic timescale of variation of the emission
as seen by the external observer is called $\delta t$ and is generally
taken to be of order (in any case, at least) $R/c\Gamma ^2$.

Henceforth, primed quantities denote quantities measured in the
comoving frame, whereas unprimed quantities refer to quantities
measured in the frame of the central engine.
Subscripted quantities refer to that quantity expressed as the power of
ten in cgs units denoted by the subscript. $R_{12}$ for example means the
radius R expressed in units of $10^{12}$ cm.

\section{X-Ray Paucity}

     A noteworthy feature of gamma ray burst spectra is the so-called
 paucity of X-rays.  Although some BATSE spectra (though apparently not
all, Preece $\it et$ $\it al.$ 1999) may be consistent with the
instantaneous synchrotron radiation spectrum of a mono-energetic
electron population, which would go as $\nu I(\nu) d\nu \propto
\nu^{4/3}d\nu $, they are often too hard to be the time integrated
spectrum of such a population, $\nu I(\nu) \propto {\nu}^
{\frac{1}{2}}$, often too hard to be the optically thin
instantaneous (i.e. thin target) emission of a shock accelerated
spectrum, $\nu I\propto \nu^{\frac{1}{2}}$, and usually too hard
to be the thick target emission spectrum of shock accelerated
electrons $\nu I(\nu) \propto $ constant. For  reviews of spectral
distribution see Pendleton {\it et al.} (1994) and Band {\it et
al.} (1993).  Popular fireball models often assume that the
primary peak of the burst radiation comes from {\it thermal}
electrons that have been accelerated to large enough energies to
synchrotron radiate at several hundred KeV, but this, in our
opinion, requires much fine tuning.  It requires that the magnetic
field value, the electron energy and the bulk Lorentz factor
always conspire to put the peak at several hundred KeV; it is
seldom below that for most GRB's. (It has been debated that the
narrow range of $\nu I_{\nu}$ peaks seen in GRB's can be
attributed to instrumental effects [Dermer, {\it et al.} 1998].
Here we are motivated to find a physical explanation.) Another
problem with optically thin synchrotron emission is that invoking
a thin target  spectrum means low radiation efficiency, unless
there is additional fine tuning. On the other hand allowing the
electrons to radiate  most of their energy in the magnetic field
("thick target emission") would make too much softer X-radiation.

Invoking optically thin inverse Compton emission (IC) for the peak
gamma rays would seemingly exacerbate the problem, because the
energy of the seed photons is less likely to be monoenergetic than the
virtual photons in the magnetic field, and even more fine tuning would be
required to ensure that the upscattered photons always peak at $\sim 300$
KeV.  [A saturated IC spectrum has far fewer free parameters (Liang 1998)
and is indeed considered below, but the location of the peak still depends
on the total number of quanta produced per unit energy.  So we will
consider such production first.]

Here we interpret the X-ray paucity as being due to self absorption.  The
simplest version of this interpretation is to identify the typical
GRB peak at $\sim 300 KeV$ as being the Wien peak of a thermal spectrum
with a temperature of $\sim 10^2/\Gamma$ KeV, where $\Gamma$ is the bulk
Lorentz factor.
 If we interpret the paucity of X-rays from GRB's as being  due
to a black-body limit, then we can infer that the temperature of the
radiation
as seen by the observer must typically be $\sim 10^2 KeV$ or more.
If the brightness temperature at the frequency of maximum
brightness temperature $\omega _m$ is indeed at the black body
limit $T_m\sim h\omega _m$, then the energy density in the frame of
the fireball, $a  T_m^4$ is limited by the condition that

\begin{equation}
 4\pi R^2 c  \Gamma ^2 aT^4 \leq 10^{51}L_{51} erg s^{-1}.
\end{equation}

The condition that the spectra typically peak at $\epsilon_{\gamma} \sim
300\epsilon_{300}$ KeV implies that

\begin{equation}
\Gamma T_m \sim 10^9 K \epsilon_{300}.
\end{equation}

This implies that

\begin{equation}
\Gamma \geq 10^{4.25}L_{51}^{-\frac{1}{2}} R_{13} \epsilon_{300}^2.
\end{equation}

Although this admits unlimited values for both $\Gamma$ and R, we note
the following simple argument that could guide a choice of both:
Liberating $\sim 10^{51}$ ergs within neutron star dimensions creates
fewer photons, with larger average energy, than dictated by GRB
observations, so some additional photon production must have taken place.
If the (photon creating) dissipation takes place at modest
$\Gamma$ (say $\Gamma\leq 10$),
then equation (3) implies that $R \leq 10^{10}$ cm for $L_{51}$ and
$\epsilon_{300}$ of unity. If the dissipation
were to take place at large $\Gamma$ and
perhaps larger R, the question would arise  as to why the combination
 of the
two always conspired to yield average photon energies of order several
hundred KeV. On the other hand,  assuming that the dissipation takes
place at modest $\Gamma$
 provides a  natural reason why the photon production should stop
once the average photon energy is comfortably below (but not much
further below) the pair production threshold: the cycle of pair
production by photons and photon production by pairs naturally
terminates at this point (e.g. Cavallo \& Rees, 1978; Blandford \&
Levinson, 1995). For example, we might associate the additional
photon production with the sharp, large angle deflections the flow
might experience while being collimated into jets by a surrounding
baryonic outflow. If thermalized at modest $\Gamma$, say at  a
dissipation radius
 $R_{dis}$, and the
flow
 expands more or less  adiabatically beyond that point, then the temperature
would from there on
scale as 1/R, while $\Gamma$ would scale as R as long as the fireball
remained baryon pure (to avoid a minimum rest mass per unit energy release)
and sufficiently optically thick (so that pairs were in thermal equilibrium
with radiation).  Thus equation (3) would continue to be satisfied as
long
as there were no further processing of the thermal photons.
Neglecting, tentatively, non-thermal pair production, the thermal pair
density would become negligible at a temperature well below $3\times 10^8$K
(Paczynski 1986; Goodman 1986),
and would thus establish a photosphere at a radius of about $10R_{dis}$,
with a $\Gamma$ of order 10 or less.
The value of $\Gamma$ could increase to  $10^3$ within a radius of
$10^{12}$ cm as long as there were no baryon loading.

The possibility of non-thermal or grey body (rather than black body)
 pair production would allow for somewhat
larger dissipation radius than given by equation (3).  This possiblity
should not alter many of the conclusions of the paper much (though see
immediately below). The basic
point
is that the modest $\Gamma$ that we invoke for the  production site of the
GRB photons seems much smaller than the asymptotic values ($\ge 300$) that
are deduced from other considerations, and the implication is that much of
the
photon production in GRB's comes prior to the acceleration of the flow
from $\Gamma = 10$ to $\Gamma = 300$

Given equation (3), with $\Gamma$ of order 10,  the characteristic
timescale of variation of
photons from such a photosphere, if of order $R_{dis}/c\Gamma ^2$,  comes
out naturally to be of order 3 milliseconds.

As has been noted many times, gamma rays observed by COMPTEL at about 20
MeV would pair produce at these radii unless the bulk  Lorentz factor
$\Gamma$ exceeded about 20.  On the other hand, beyond the photosphere,
the
bulk Lorentz factor of a baryon poor wind can grow in proportion to R, so
that the condition $\Gamma \geq 20$ is attained not much beyond the
photosphere.  We contend that any COMPTEL gamma ray emitted at $R\geq 3
\times 10^{10}$ cm at $\Gamma \geq  30$ would not be subject to pair
production and could in principle
 reproduce time variations as short as 10
ms.

In the particularly well studied burst GRB 930131, COMPTEL detected gamma
rays out to 20 MeV or so with a time profile very similar to that of the
BATSE profile, and we suggest that this be interpreted as both energy
ranges reflecting the intrinsic time profile of the energy release.
On the other hand, the EGRET gamma rays at 30 MeV, of which there were
several, typically arrived several seconds after
the  BATSE and COMPTEL peaks,  and this is consistent with the hypothesis
that the shortest time profile
for the EGRET gamma rays was of order seconds.

\section{High Energy Gamma Rays: Production}

Although  it is tempting if only for simplicity to  model GRB as coming
from a particular
characteristic radius, the following argument suggests that
a broad range of emission radii is no less natural.
As long as the fireball is not baryon loaded and not complicated by
non-spherical expansion, the bulk Lorentz factor
increases linearly with R. The proper time in the frame of the fireball
is then described by

\begin{equation}
dt' =  dt/\Gamma \propto  dR/R.
\end{equation}

It follows that the proper time evolves only logarithmically with
R.   The duration of shock  activity (e.g. due to very clumpy baryon
contamination, fluctuations in $\Gamma$ at launch, etc.) in the
proper frame $\Delta t^{\prime}$, which is likely to be of order
the mean time $t^{\prime}$ around which such activity is centered, 
thus persists over a large range of
radii R. That is, the range of radii over which there is shock
activity, $\Delta R$, is given by $\Delta lnR \sim ln(\frac
{R}{R_0})$, where $R_0$ denotes the beginning of the acceleration zone
and might be close to $R_{dis}$. If $\frac {R}{R_0} \gg 1$, then shocks
could persist over many decades of R, particularly shocks associated with
the launch of the fireball, which can be delayed or "dragged out" to $R 
\gg R_o$ by the
time dilation. Previous discussions of internal
shocks occuring at some characteristic radius (Levinson \& Eichler
1993, Eichler 1994, Rees \& Meszaros 1994) have perhaps not
emphasized this point. (We agree, however, that shocks resulting purely
from fluctuations in the saturation radius, at which $\Gamma$ approaches
its asymptotic value, are likely to form near the mean value of that
radius if the launch conditions are otherwise identical and if the
baryonic contamination is not too clumpy.) We contend
that there is
no reason to expect all of the gamma rays from GRB's to come from the same
 radius. We therefore consider the evaluation of
gammaspheric radii as a function of gamma ray energy.

In order  to emit gamma rays at energy E, the fireball must be
able to accelerate the particles to at least energy E, and the gamma rays
must then be able to escape.

The first condition is expressed as

\begin{equation}
\gamma _m \sigma _T U^{\prime} /m_e c = \eta eB^{\prime} /\gamma _m m_e c
\end{equation}
where $\gamma _m $ is the maximum Lorentz factor in the frame of the
fluid that is achievable by the acceleration,  $U^{\prime}$ is the energy
density of the radiation field in the frame    of the fireball fluid
element, $B^{\prime}$ is the field strength in the fluid frame in cgs
units and $\eta$ is the acceleration rate in
units of the electron gyrofrequency. For any reasonable acceleration
mechanism, $\eta$ should be less than unity and for shock acceleration a
value of order 1 to 10 percent seems like a reasonable guess. Writing
$U^{\prime}$
as $10^{51}L_{51} erg s^{-1}/4\pi R^2 c\Gamma^2$ equation (5) implies that

\begin{equation}
\gamma _m/ \Gamma  = 0.5 \eta ^\frac{1}{2} R_{12}
B^{\prime \frac{1}{2}} L_{51}^{-\frac{1}{2}}.
\end{equation}

For $L_{51}=1$, and assuming that the Poynting flux is limited by
$10^{51}L_{51}$ ergs per $4\pi $ steradians, the last equation reduces to

\begin{equation}
\gamma _m \Gamma = 6 \times 10^3 \eta ^\frac{1}{2}
\Gamma ^\frac{3}{2} R_{12}^\frac{1}{2}.
\label{eq:gG}
\end{equation}
Note that if $B=\Gamma B^{\prime}$ scales as $1/R$, then the maximum
energy in the frame of
the fireball to which electrons can be accelerated increases as
$R^{1/2}\Gamma^{1/2}$, as seen from eq. (6).
This implies  that the ability of the fireball to accelerate
electrons
to extremely high energies increases sharply with radius. At small
radii,
\begin{equation}
R_{12} \leq 2 \eta^{-\frac{1}{2}} B^{\prime -\frac{1}{2}}
\end{equation}
a shock cannot accelerate electrons to a high enough
energy to pair
produce via inverse Compton scattering of thermal
photons,.
(i.e. to a Lorentz factor of $\gamma_{m} \geq \Gamma$,
which would be needed given
 that the photons near the spectral peak
have energies less than $m_ec^2/\Gamma$ in the fluid
frame).

At  a radius of order $ 10^{12}$ cm, with $\Gamma \geq
40$, equation (7) admits TeV gamma rays which could arrive
within a fraction of a second if they escape freely.

If non-thermal pairs are made at larger radii than given in equation (8),
they could conceivably Compton scatter the thermal photons that
were made at smaller radii. If the thermal photons constitute
most of the burst's photons and most (or much of) its
energy, further scattering by pairs downstream
cannot greatly alter the parameters of the thermal component.

\section{High
Energy Gamma Rays: Escape}

In order to compute the pair production opacity contributed by
the beamed photons and the associated gammaspheric radii, we
consider a conical beam of emission with an opening
semiangle $\theta_b$, and a power law spectrum which is taken to be
constant with radius: $n(\epsilon_s)=n_o\epsilon_s^{-\alpha}$;
$\epsilon_{min}<\epsilon<\epsilon_{max}$ where $n_o$ is expressed
in terms of the apparent luminosity, $L$, as
\begin{equation}
n_o=\frac{L}{4\pi^2 m_ec^3 r^2}(1-\mu_b^2)^{-1}
G(\epsilon_{max},\epsilon_{min})\simeq \frac{L\Gamma^2}{4\pi^2 m_ec^3 r^2}
G(\epsilon_{max},\epsilon_{min}),
\label{eq:no}
\end{equation}
where $\mu_b=\cos\theta_b$, $G(\epsilon_{max},\epsilon_{min})=
[\ln(\epsilon_{max}/\epsilon_{min})]^{-1}$
for $\alpha=2$, and  $G(\epsilon_{max},\epsilon_{min})=(2-\alpha)/
(\epsilon_{max}^{2-\alpha}-\epsilon_{min}^{2-\alpha})$ for
$\alpha\neq2$.
The pair production opacity at energy $\epsilon_{\gamma}$ (measured
in units of $mc^2$) can then be expressed as
\begin{equation}
\kappa_p(\epsilon_{\gamma})=2\pi\int_{\epsilon_{min}}^
{\epsilon_{max}}d\epsilon_s
\int_{\mu_b}^{\mu_{max}}d\mu\{(1-\mu)n(\epsilon_s)\sigma_p(\beta)\}.
\label{eq:kappa}
\end{equation}
Here $\mu_{max}=$ max$\{\mu_b; 1-2/\epsilon_s\epsilon_{\gamma}\}$, and
$\sigma_P(\beta)$ is the pair-production cross section, where
$\beta$ is the speed of the electron
and the positron in the center of momentum frame, and is given by
\begin{equation}
(1-\beta^2)=\frac{2}{(1-\mu)\epsilon_{\gamma}\epsilon_s}.
\end{equation}
The threshold energy, obtained for $\beta=0$ and $\mu=\mu_b$, now reads:
$\epsilon_{thrs}=2/(1-\mu_b)\epsilon_{\gamma}\simeq 4\Gamma^2/
\epsilon_{\gamma}$.
Substituting eq. (\ref{eq:no}) into eq. (\ref{eq:kappa}) yields
\begin{equation}
\kappa_p=\frac{3\sigma_T L}{16\pi m_ec^3 r^2}\frac{1}{2^{2\alpha}
\Gamma^{2\alpha}}G(\epsilon_{max},\epsilon_{min})A(\alpha,\beta_{max})
\epsilon_{\gamma}^{\alpha-1},
\end{equation}
where
\begin{equation}
A(\alpha,\beta_{max})=\frac{1}{\alpha+1}\int_{0}^{\beta_{max}}
d\beta \beta(1-\beta^2)^{\alpha-1}\left[(3-\beta^4)\ln\left(
\frac{1+\beta}{1-\beta}\right)-2\beta(2-\beta^2)\right].
\end{equation}
The function $A$ can be evaluated numerically.  It vanishes for
energies below the threshold energy at which $\beta_{max}=0$.  A
plot of $A$ versus $\alpha$ for $\beta_{max}=1$ (which is a good
approximation well above the threshold energy) is given in
Blandford \& Levinson (1995).  Numerically we find $A=0.2$ for a
photon index $\alpha=2$.

Now, the gamma-spheric radius, $r_{\gamma}(\epsilon_{\gamma})$, is
defined implicitly by the equation
\begin{equation}
\int_{r_{\gamma}}^{\infty}dr \kappa_p(r,\epsilon_{\gamma})=1.
\end{equation}
Using the above expression for the opacity, one obtains

\begin{equation}
r_{\gamma}(\epsilon_{\gamma})=1.6\times10^{21}L_{51}(2\Gamma)^{-2\alpha}
G(\epsilon_{max},\epsilon_{min})A(\alpha,\beta_{max})\epsilon_{\gamma}^
{(\alpha-1)}\ \ \ \ \ {\rm cm},
\end{equation}
up to a numerical factor that depends on the radial profile of the
Lorentz factor $\Gamma$.  Adopting, for illustration, $\alpha=2$,
$\epsilon_{min}=2$, $\epsilon_{max}=2\times10^3$, yields a threshold
energy of 10$\Gamma_2^2$ MeV, a gammaspheric radius
\begin{equation}
r_{\gamma}(\epsilon_{\gamma})=2.8\times10^{10}\frac{L_{51}}{\Gamma_2^{4}}
\epsilon_{\gamma}\ \ \ \ \ {\rm cm}
\end{equation}
above the threshold energy, and a corresponding variability time
\begin{equation}
\delta
t=\frac{r_{\gamma}}{2c\Gamma^2}=5\times10^{-5}\frac{L_{51}}{\Gamma_2^{6}}
\epsilon_{\gamma}\ \ \ \ \ s.
\end{equation}

These last equations admit rather high energy gamma rays emerging
from rather compact regions.  Even the GeV gamma rays observed by EGRET
can in principle emerge from within $10^{12}$ cm if the bulk Lorentz
factor is 300 or more.

The above results may be considerably altered in the presence of
additional radiation component.  For instance, large angle
scattering of the beamed photons by ambient gas surrounding the
fireball, e.g., a confining, baryon-rich outflow (Nakamura 1998;
MacFadyen \& Woosley 1998; Eichler \& Levinson 1999), can produce
an unbeamed component that will dominate the pair production
opacity within the baryon-poor jet
even if its luminosity is only a small fraction of the total. 
It may then be that high energy gamma rays can escape only from parts of
the baryon-poor jet that have run ahead of surrounding baryons.

\section{Comparison of Constraints on Baryon Contamination}

The cooling of the plasma to $2\times 10^8$ K, so that the
Boltzman factor for pairs becomes miniscule, is a necessary
condition for small photospheres, but it is not a sufficient one.
Electrons may be entrained from the sides (e.g. via neutron
seepage) and non-thermal pairs may be produced under some
conditions.  Let us then consider the limits on the net baryon
number within a jet that features a compact photosphere.

Writing

\begin{equation}
\dot M = 4\pi m_p n R^2 \beta c
\end{equation}
and using $\sigma_T n R = \tau \Gamma^2$, where $\tau$ is optical
depth,  and letting $\tau$ be unity at the photospheric radius
$R_{ph}$, we easily obtain

\begin{equation}
\Gamma_a \Gamma^2_{ph} \ge 2.5 \times 10^{5} L_{51}R^{-1}_{ph13}
\end{equation}
where $\Gamma_a$ is defined as $L/\dot Mc^2$.

If the photosphere were to occur at  $\Gamma_{ph} \sim 10$, and
$R_{ph}$ of order $10^{10}$ cm,  then $\Gamma_a$ would have to be
enormous, of order  $10^6$, and might be convincing evidence that
an event horizon enforces baryon purity on field lines that thread
it. On theother hand, the compactness constraint that motivates
our model does not imply that the photons that have been produced
within $10^{10}$ cm necessarily make their last scattering there,
it merely implies that they are not swamped  by a much larger
number of photons made at larger radii, nor drained of their
energy by mass loading (which would terminate the linear increase
of $\Gamma$ with R).  They may be scattered by material that is
both dynamically and thermodynamically passive; the timescale and
spectrum could remain the same.

We may combine equations (3), (19) and the obvious condition that
$\Gamma_a \geq \Gamma_{ph}$, to obtain that at $R_{ph}\sim
10^{11}$cm, $\Gamma_a $ must exceed $200$. If the photosphere were
constrained by some other consideration  to be larger than this,
then equations (3) and (19) similarly constrain $\Gamma_{ph}$ to
be greater as well.

\section{Conclusions}

We have suggested that a significant fraction  of the photons from
gamma ray bursts are emitted at a radius of about $ 10^{10}$ cm or
less , where  the bulk Lorentz factor is  modest. This allows the
burst entropy to be established at modest $\Gamma$, so that the
peak energy in the observer's frame can be tied directly to
$m_ec^2$ as oppsed to a small fraction of $m_ec^2$ to be
multiplied by a large $\Gamma$. It also provides  a natural
account of why  GRBs have time substructure of milliseconds.

The photosphere (surface of last scattering) may be larger than the
radius of emission,  but whether
inequality (3) is satisfied is independent of R  as long as $\Gamma$
remains proportional to R.
Thus, the fulfillment of (3) is unaffected by passive
scattering material.

Higher energy, non-thermal gamma
rays may be produced even beyond the photosphere of lower energy ones,
but we found in section IV that  even GeV gamma rays could escape from
within $10^{12}$ cm for $\Gamma$ of order 300.

TeV gamma rays could,  in principle, escape if produced within $10^{12}$
to $10^{13}$ cm
 for $\Gamma$ of order $10^3$ and would be prompt.
Thus, it seems worthwhile to run experiments that could detect
prompt, ultrahigh, non-thermal gamma rays from an unannounced
direction. The MILAGRO project is such a detector, though the
universe may be opaque to the photons ($\epsilon_\gamma \geq 1$
TeV) to which it is sensitive. A similar experiment with a lower
energy threshold - say $10^2$ GeV - might detect prompt UHE
emission from a sizeable fraction of GRB's. Detection of prompt gamma
rays above 100 GeV  might be the best diagnostic of the  bulk
Lorentz factors of GRB fireballs.

     To conclude, the radius of emission  and radius of last scattering of
thermal gamma rays in GRB's may be different, but both may occur well
within $10^{12}$ cm.  Non-thermal, high energy gamma rays may be produced
at and escape from somewhat larger radii than the thermal ones, but they
too can emerge from compact regions provided that the bulk Lorentz factor
is sufficiently high. Soft echos of primary gamma radiation scattered off
baryonic matter might provide (via their time scales) limits on the
photospheric radius for the bulk of the gamma rays.

     We acknowledge support from the Israel Science Foundation.
We thank Dr. E. Waxman for useful discussions.

\break

\end{document}